# Electron-Phonon Coupling and Quantum Correction to Topological Magnetoconductivity in Bi₂GeTe₄

Niraj Kumar Singh,[1] Divya Rawat,[1] Dibyendu Dey,[2] Anna Elsukova,[3] Per O. Å. Persson,[3] Per Eklund,[3] A. Taraphder[4] and Ajay Soni[1]*

[1]*School of Basic Sciences, Indian Institute of Technology Mandi, Mandi 175075, Himachal Pradesh, India*

[2]*Department of Physics and Astronomy, University of Maine, Maine 04469, USA*

[3]*Thin Film Physics Division, Department of Physics Chemistry and Biology (IFM), Linköping University, Linköping, SE-58183, Sweden*

[4]*Centre for Theoretical Studies, Department of Physics, Indian Institute of Technology Kharagpur, Kharagpur 721302, India*

*Author to whom correspondence should be addressed: ajay@iitmandi.ac.in

We report on structure, vibrational properties and weak-antilocalization-(WAL-) induced quantum correction to magnetoconductivity in single crystal Bi₂GeTe₄. Surface band structure calculations show a single Dirac cone corresponding to topological surface states in Bi₂GeTe₄. An estimated phase coherence length, $l_\phi$ ~ 143 nm and prefactor $\alpha$ ~ - 1.54 from Hikami-Larkin-Nagaoka fitting of magnetoconductivity describe the quantum correction to conductivity. An anomalous temperature dependence of $A_{1g}$ Raman modes confirms enhanced electron-phonon interactions. Our results establish the involvement of vibrations of Bi-Te with existence of topological surface states and WAL in Bi₂GeTe₄.

Keywords: Layered chalcogenides, Bi₂GeTe₄, Raman spectroscopy, Phonon modes, Topological insulators

Topological insulators (*TIs*) are a class of novel quantum materials characterized by conducting spin-momentum locked surface states and insulating bulk. The gap-less surface states constitute an odd number of Dirac cones at the Fermi surface due to relativistic effects such as spin orbit coupling (*SOC*) and are topologically protected by time-reversal-symmetry. [1-5] The Dirac cone hosts spin helicity in the momentum space leading to a $\pi$ Berry phase, which can limit the backscattering of surface electrons for enhanced conductivity. Because of the interference of the time-reversed electron paths, a quantum correction to the conductivity is observed in magnetotransport studies. This effect is called weak antilocalization (WAL), a common signature of topological insulators originated from the *SOC*. The 3D topological insulator states are commonly defined by set of four topological invariants given as $[v_0; v_1 v_2 v_3]$, where $v_0$ represents strong invariant while $v_1$, $v_2$, and $v_3$ are the weak topological invariants.[4] The topological invariants distinguish the strong (*odd number of Dirac cones*) and weak (*even number of Dirac cones*) topological insulators. In the presence of disorder, the strong *TIs* are robust while the weak *TIs* behave like a band insulator. *TIs* are of great interest because of their applicability in quantum computing, spintronic devices and axion electrodynamics.[6,7] *TIs* were first predicted theoretically and later observed experimentally in 2D HgTe and further in 3D $(Bi/Sb)_2(Te/Se)_3$.[8,9] Bismuth chalcogenides have been widely studied for their topological properties mainly due to the fact that their thin films are relatively easy to grow and the Dirac point can be accessed at nominal doping levels.[6,10] Recently, ternary layered chalcogenide materials such as $Bi_2(Te/Se)_2(Te/Se/S)$, $Bi_2GeTe_4$, $PbBi_4Te_7$ have been studied extensively for topological properties.[11] Among the various ternary layered chalcogenide *TIs*, $Bi_2GeTe_4$ is a newly discovered strong *TI* with non-zero weak topological invariants as [1;111] class.[5,12] Angle-resolved photoemission spectroscopy results of $Bi_2GeTe_4$ showed the existence of an isolated

Dirac cone with bulk band gap of ~ 0.18 eV.[5,13] $Bi_2GeTe_4$ has three septuplet atomic slabs in the unit cell arranged in the sequence $Te^I$-$Bi$-$Te^{II}$-$Ge$-$Te^{II}$-$Bi$-$Te^I$, where $^I$ and $^{II}$ denotes different chemical bonding associated with Te.[14] The $Bi$-$Te^{II}$ and $Bi$-$Te^I$ bonds are predominantly covalent in nature with minor ionic character, while $Te^I$-$Te^I$ bonds are weak van der Waals forces (vdW). Since Ge occupies the center of mass, it remains inactive in the lattice dynamics. Taking into consideration that Raman spectroscopic studies performed on layered chalcogenides such as $Bi_2Te_3$ and $Bi_2Se_3$, have produced in-depth understanding of the phonon dynamics, therefore it is a pressing need to probe the behavior of phonons in $Bi_2GeTe_4$.[15] With the limited literature on the structural, thermoelectric and band structure calculations, $Bi_2GeTe_4$ has not been explored for its vibrational properties and magnetotransport properties which can provide evidences of the topological surface states.[14,16-18]

In this letter, we report on the structural, vibrational, electronic and magnetotransport studies supported by the theoretical calculations for single crystalline $Bi_2GeTe_4$. The experimental details are provided in the supplemental material.[19] Electron microscopy reveal a hierarchical layered crystalline nature ranging from atomic to higher levels. A sharp rise in magnetotransport, near zero magnetic field, specifies the quantum correction to the conductivity arising due to *WAL*. The analysis of the anomalous temperature dependence of out-of plane $A_g$ symmetry Raman modes, involving Bi and Te stretching, suggest electron-phonon coupling in single crystalline $Bi_2GeTe_4$.

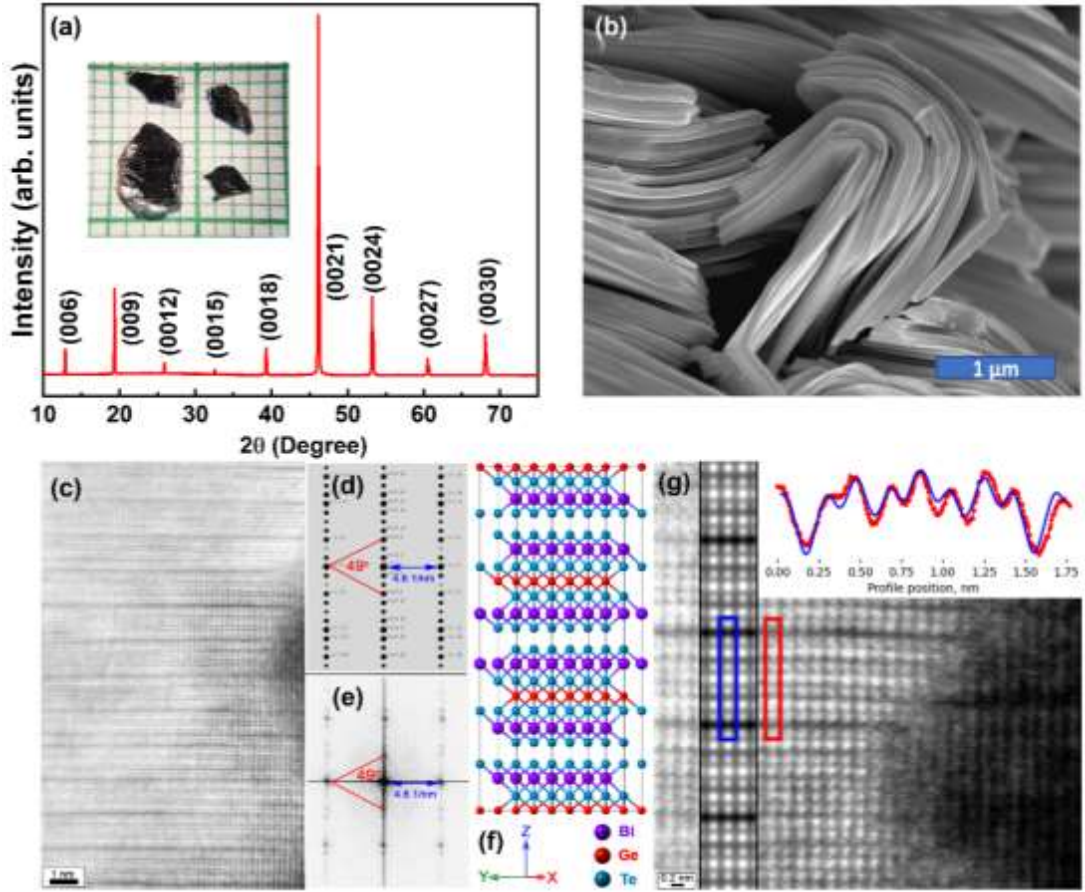

FIG. 1. (a) X ray diffraction pattern for single crystal $Bi_2GeTe_4$ (inset: photograph of the crystals). (b) Cross-sectional FESEM image of fractured crystal showcasing the stacked layers. (c) Cross-sectional STEM image, (d) simulated electron diffraction and (e) corresponding FFT pattern of the (f) $Bi_2GeTe_4$ crystal viewed in [210] direction. (g) HR-STEM image acquired at higher magnification. The overlay shows the simulated HR- STEM image of the $Bi_2GeTe_4$ crystal viewed in [210] direction while the inset presents averaged intensity line profiles obtained from the experimental (red rectangle) and simulated (blue rectangle) HR-STEM images.

The X-ray diffraction pattern of phase pure $Bi_2GeTe_4$ (Fig. 1 (a)) shows that it crystallizes into rhombohedral crystal geometry ($R\bar{3}m$, 166). The inset in Fig. 1(a) shows a photograph of the as-grown crystals of size in the range of 3 - 6 mm. A Rietveld refinement of $Bi_2GeTe_4$ powder XRD patterns (Fig. S1) yields lattice parameters of *a = b = 4.33 Å* and *c = 41.35 Å* and the unit cell volume is *670.8 Å³*.[18,19] An FESEM image of the cross-section of a freshly cleaved flake (Fig. 1 (b)), shows the layered nature of $Bi_2GeTe_4$ with no signs of secondary phases in elemental mapping (Fig. S2). [19] Lattice resolved STEM imaging (Fig. 1 (c)), shows the periodicity of vdW gaps along the *c*-axis with repetition of atomic septuplets in the sequence *Te^I-Bi-Te^II-Ge-Te^II-Bi-Te^I*.[14] The brightness contrast in the images reveals evenly spaced septuplet (~1.3 nm) structures. The diffraction pattern (Fig. 1 (d)) closely matches the Fast Fourier Transform (FFT) pattern (Fig. 1 (e)) simulated for $Bi_2GeTe_4$ crystal viewed in [210] direction (Fig. 1 (f)).[20,21] The formation of the structure was further confirmed by a simulation of the STEM image and overlaying it with experimental one (Fig. 1 (g)). The intensity distribution profile in the experimental image matches closely with the simulated STEM image (Fig. 1 (g), top-left inset). Overall, the XRD and TEM results confirm the single crystalline nature of the as-grown $Bi_2GeTe_4$.

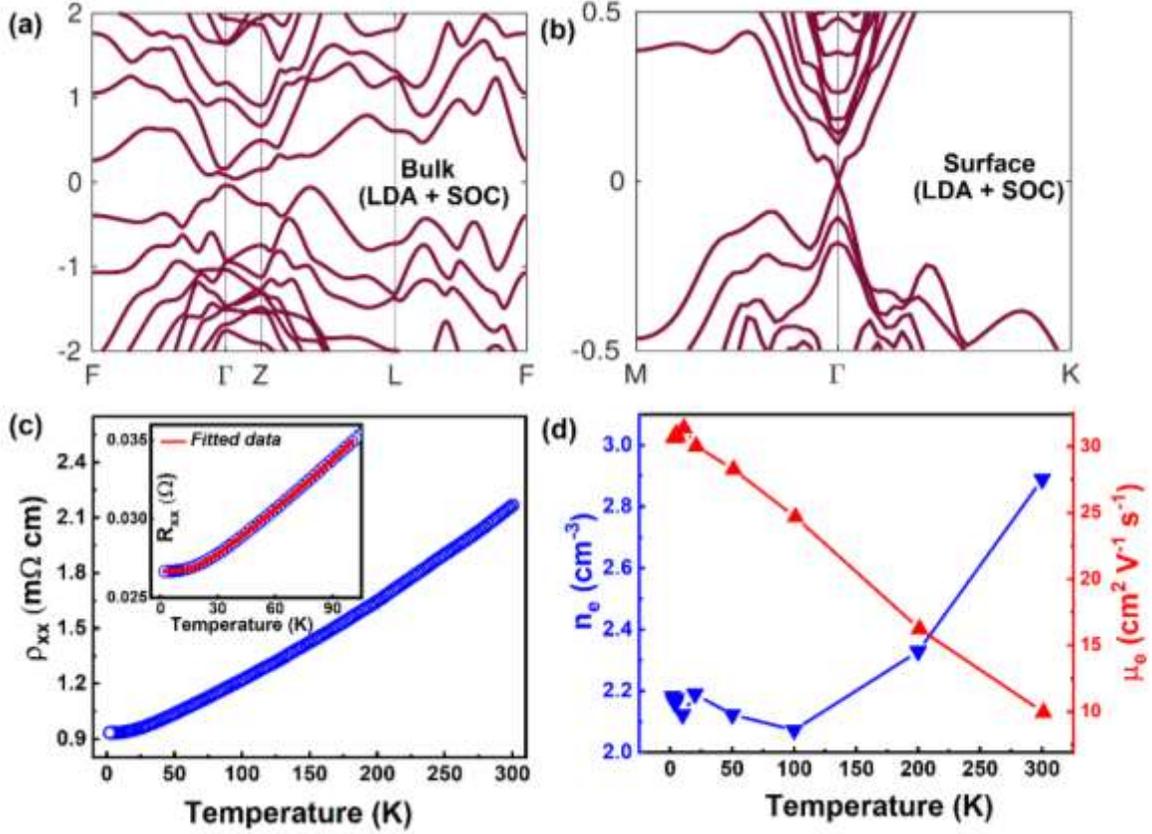

FIG. 2. (a) Bulk electronic band structure of $Bi_2GeTe_4$ within LDA + SOC. The high symmetry points of the rhombohedral Brillouin zone (BZ): $\Gamma = (0,0,0)$, $Z = (½,½,½)$, $L = (½,0,0)$, $F = (½,½,0)$ have been used in the band structure plots. (b) surface (111) band structure is calculated (within LDA+SOC) along the high symmetry points of the hexagonal BZ: $\Gamma = (0,0,0)$, $M = (½,0,0)$, $K = (⅓,⅓,0)$.[22] (c) The temperature dependence of longitudinal resistivity, $\rho_{xx}$, and fitting of $R_{xx}$ (Inset), and (d) the calculated carrier densities ($n_e$) and mobilities ($\mu_e$) at various temperatures. Note: The $n_e$ has the order of $10^{20}$ cm$^{-3}$.

The electronic band structures of bulk $Bi_2GeTe_4$ with SOC (Fig. 2 (a)) and without SOC (Fig. S3) have finite band gap.[19,23-26] Thus, SOC plays a crucial role in predicting the correct dispersion where the band gap changes from ~ 0.38 eV (LDA) to ~ 0.1 eV (LDA+SOC), which is close to the reported bulk band gap ~ 0.18 eV.[5,13] When the (111) surface is cleaved from the bulk, the crystal symmetry changes from rhombohedral to hexagonal, and a single gapless surface band appears at the $\Gamma$ point (Fig. 1 (d)). Observation of Dirac point indicates that $Bi_2GeTe_4$ belongs to the $Z_2 = -1$ topological class, which is consistent with the earlier results.[5]

The temperature dependence of longitudinal resistivity ($\rho_{xx}$) shows a degenerate semiconductor behavior of the $Bi_2GeTe_4$, (Fig. 2 (c)). The residual resistivity ratio, RRR ~ 2.33, determined from $\rho_{xx}$ values at 300 K (~ 2.17 mΩ-cm) and 2 K (~ 0.93 mΩ-cm), indicates a relatively good crystal quality. Layered chalcogenides are known for anisotropic carrier relaxation times arising due to vdW gaps along the *c*-axis, and anharmonic phonons.[27] Due to multiple scattering mechanisms and interference from the lattice, layered chalcogenides deviate from the Matthiessen's rule.[28] Therefore, $R_{xx}(T)$ is fitted using a phenomenological model: $R_{xx} = R_0 + \alpha_R e^{-\theta/T} + \beta_R T^2$; where $R_0$ is residual resistance, θ represents effective phonon frequency ($\omega = \frac{\kappa_B \theta}{h}$) and the $\alpha_R$ and $\beta_R$ appears for phonon scattering and electron-electron scattering, respectively (inset Fig. 2 (c)). Exponential term corresponds to the intervalley scattering which involves either zone-boundary acoustic phonons or low-energy optical phonons. The fitting of $R_{xx}(T)$ yields $\alpha_R$ ~ $7.8 \times 10^{-3}$ Ω, $\beta_R$ ~ $4.45 \times 10^{-7}$ Ω-K$^{-2}$ and effective phonon frequency ω ~ $1.38 \times 10^{13}$ rad/s. This smaller value of $\beta_R$ suggests that electron-electron scattering is negligible in $Bi_2GeTe_4$.[29] The relation $\rho_{xy}(B) = (\rho_{xy}(+B) - \rho_{xy}(-B))/2$ has been used to anti-symmetrize the raw Hall data in order to compensate for the residual offset voltages due to the geometrical misalignment.[6,30] The temperature dependence of Hall resistivity ($\rho_{xy}$) reveals the electron-dominated charge transport and the estimated carrier density ($n_e$) increases with the increasing temperature and attains a value of ~ $2.9 \times 10^{20}$ cm$^{-3}$ at 300 K, (Fig. 2 (d)). Additionally, the calculated carrier mobility ($\mu_e$) decreases with temperature and has a value of ~ 10 cm$^2$V$^{-1}$s$^{-1}$ at 300 K.

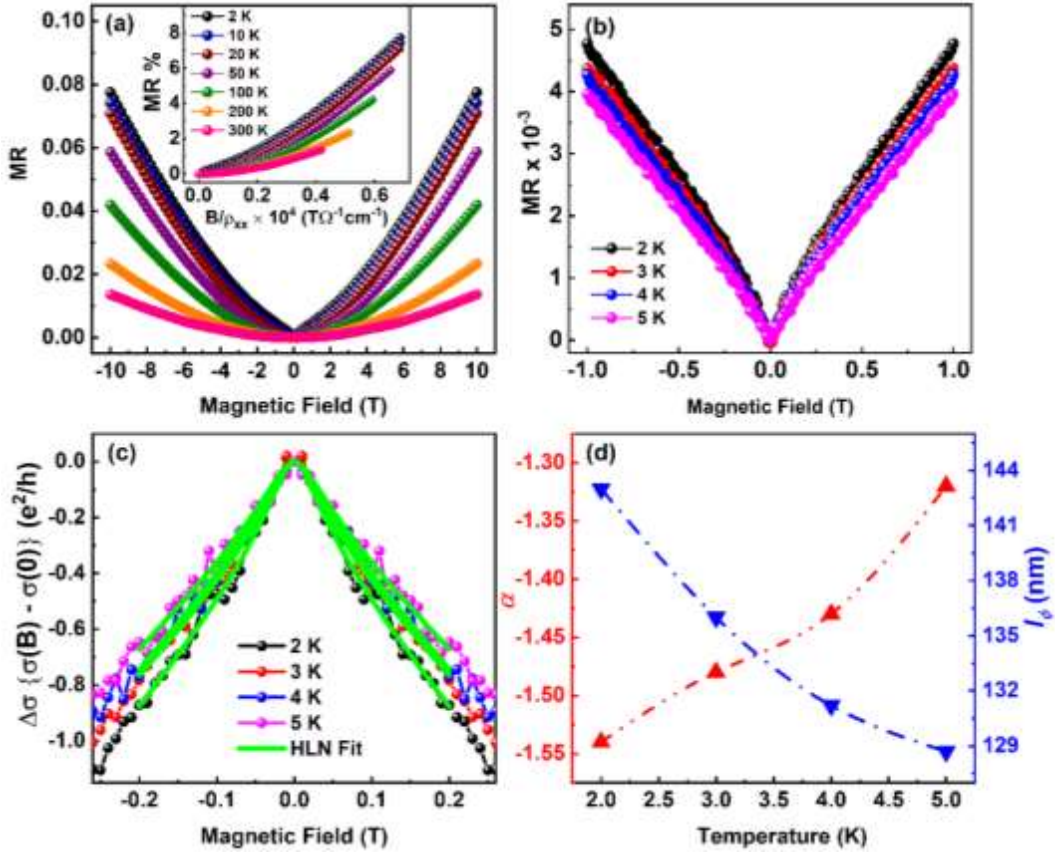

FIG. 3. The magnetoresistivity (MR% = ($\Delta\rho/\rho$) x100) of $Bi_2GeTe_4$ at various temperatures in the magnetic field range of (a) ± 10 T, and (b) ± 1T showing the characteristic cusp at zero magnetic field. (c) The low temperature HLN fitting of the magnetoconductivity data and (d) the calculated prefactor ($\alpha$) and phase coherence length ($l_\varphi$).

The magnetoresistivity (MR) is given by $MR = \frac{\rho_{xx}(B) - \rho_{xx}(0)}{\rho_{xx}(0)}$, where $\rho_{xx}(B)$ and $\rho_{xx}(0)$ are the resistivity values at applied magnetic field $B$ and zero field, respectively. The measured MR, $\rho_{xx}(B) = (\rho_{xx}(+B) + \rho_{xx}(-B))/2$, after symmetrization, is shown in Fig. 3 (a). With increasing temperature, the *MR* decreases from ~ 0.08 (at 300 K) to ~ 0.015 (at 2 K). As per the Kohler's rule, with a single dominant scattering mechanism, the MR is expected to be a universal function of $B/\rho_0$.[31] The inset of Fig. 3 (a) shows the Kohler scaling and the temperature dependence of the MR vs $B/\rho_0$ plot indicates the multiple scattering mechanisms the sample. The low field MR plot at lower temperatures show a cusp-like behavior (Fig. 3 (b)) arising because of WAL in the sample. Taking into account, the effect of classical MR, the magnetoconductivity ($\Delta\sigma$) due to WAL can be fitted with Hikami-Larkin-Nagaoka (HLN) equation given as:

$$\Delta\sigma = \sigma(B) - \sigma(0) = -\frac{\alpha e^2}{2\pi^2\hbar}\left[ln\left(\frac{\hbar}{4el_\phi^2 B}\right) - \Psi\left(\frac{1}{2} + \frac{\hbar}{4el_\phi^2 B}\right)\right] + cB^2 \qquad (1)$$

where $\Psi$ represents digamma function, $l_\phi$ is phase coherence length and $\alpha$ is a dimensionless fitting parameter.[32] The HLN model has been applied to several bulk layered chalcogenide materials, mainly $Bi_2Te_3$ and $Bi_2Se_3$, in order to explain the quantum correction of magnetoconductivity arising due to WAL, near zero field.[6,33-36] The low field ($\pm$ 0.2 T) magnetoconductivity data is fitted with modified HLN equation and shown in Fig. 3 (c) and the extracted values of $\alpha$ and $l_\phi$ are presented in Fig. 3 (d). Generally, $\alpha$ has a theoretical value of -0.5 for 3D systems with 2D surface states for a single topological surface channel.[36] In *TIs*, ideally the bulk electronic structure is gapped while the surface is essentially gapless, however, in real *TIs*, the bulk is not entirely insulating and contributes to the conductivity of the sample.[33] Here, $\alpha$ varies from -1.54 (at 2 K) to -1.32 (at 5 K), indicating the possible coupling of bulk and surface conducting states. This coupled surface and bulk conductivity leads to deviation in the theoretical value of $\alpha$. The value of $l_\phi$ decreases with increasing temperature from ~ 143 nm (at 2 K) to ~ 129 nm (at 5 K), which indicates increase in electron-phonon interaction, since $l_\phi$ is determined by inelastic scattering.

The evident involvement of lattice with WAL in magneto transport requires further exploration of interaction of phonons with charge carriers. For this, a detailed Raman spectroscopic analysis has been performed.[15,37] The primitive cell of $Bi_2GeTe_4$ have 7 atoms, which corresponds to 21 vibrational modes with 18 optically active and 3 acoustic modes. From the group theoretical calculations, the irreducible representations corresponding to optical modes can be written as $\Gamma = 3A_{1g} + 3A_{2u} + 3E_u + 3E_g$, where E modes are doubly degenerated.[38] Among all optically active vibrational modes only six are Raman active including $A_{1g}$ and $E_g$ symmetry while six are Infra-red active having $A_{2u}$ and $E_u$ symmetry. Figure 4 (a) shows the baseline corrected normalized Raman spectra recorded at 300 K and 4 K, where five out of six theoretically predicted Raman active modes are clearly observed at room temperature.

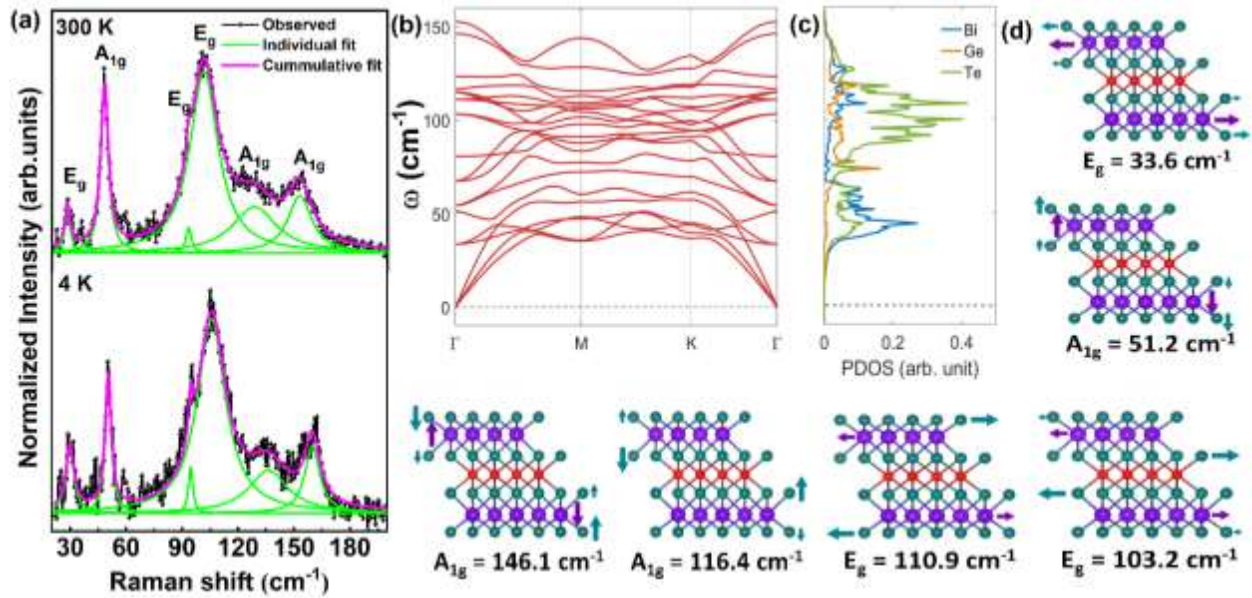

FIG. 4. (a) Normalized and fitted Raman spectra at 300 K and 4 K, (b) phonon dispersion, (c) atom projected PDOS and (d) Schematics of the Raman active modes for $Bi_2GeTe_4$.

To understand the origin and symmetry associated with the observed Raman active modes, we have performed theoretical calculations and analyzed phonon dispersion and phonon density of states (PDOS) as represented in Fig. 4 (b-c).[39,40] The phonon frequency ($\omega$) of all the Raman modes, estimated theoretically as well as observed experimentally, are tabulated in Table S3.[19] The symmetry of the modes observed at ~ 28.2 cm⁻¹, ~ 48.3 cm⁻¹, ~ 101.3 cm⁻¹, ~ 129 cm⁻¹ and ~ 152 cm⁻¹ have been assigned based on our results of calculations.[19] Atom-projected PDOS reveals that high-ω Raman active modes (>75 cm⁻¹) have major contribution arising from the vibration of Te atoms while for low-ω modes (<60 cm⁻¹) involve the vibration of both Bi and Te atoms. Here, the $E_g \sim 28.2$ cm⁻¹ and $A_{1g} \sim 48.3$ cm⁻¹ are interlayer modes. The modes at $E_g \sim 102$ cm⁻¹ and at $A_{1g} \sim 152$ cm⁻¹ corresponds to symmetric bending of the Ge-Te and asymmetric stretching of the Bi-Te, respectively.[38] The schematic of atomic vibrations associated with the various Raman modes are shown in Fig. 4 (d).

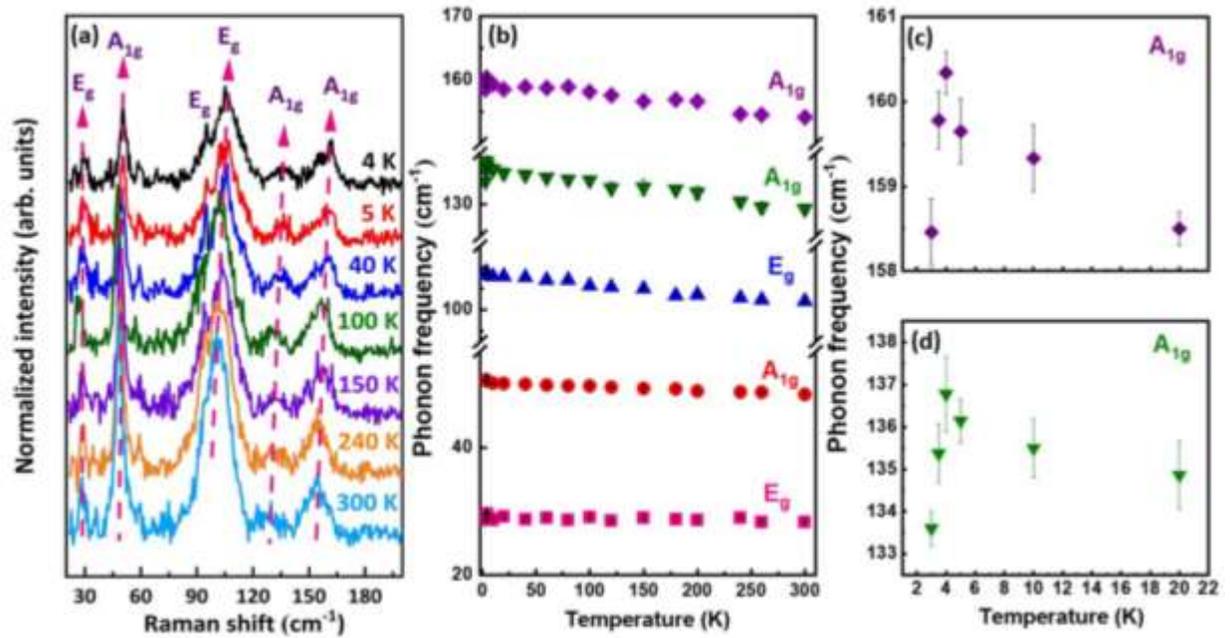

FIG. 5 Temperature dependence of (a) Raman spectra, (b) Phonon frequency, and (c-d) Anomalous response, below 10 K, of phonon frequency in $A_{1g}$ mode involving stretching of Bi and Te atoms.

The temperature dependent Raman spectra are shown in Fig. 5 (a), where all the Raman active modes are highly intense and a systematic hardening of the phonon frequencies has been observed with lowering the temperature (Fig. 5 (b)). The temperature dependence of phonon frequency is fitted with $\omega = \omega_0 + \chi T$, where $\omega_o$ is the vibrational frequency at 0 K and $\chi$ is the first order temperature coefficient (Fig. S4).[19,41] The slope, (estimated $\chi$ values), of the temperature dependence of phonon-frequency are found to be -0.006 (cm$^{-1}$K$^{-1}$), -0.021 (cm$^{-1}$K$^{-1}$), -0.018 (cm$^{-1}$K$^{-1}$) and -0.014 (cm$^{-1}$K$^{-1}$) for $A_{1g}$ (~ 48 cm$^{-1}$), $A_{1g}$ (~ 129 cm$^{-1}$), $A_{1g}$ (~ 152 cm$^{-1}$) and $E_g$ (102 cm$^{-1}$), respectively. Here, the higher $\chi$ value corresponds to stronger coupling of phonons with electrons.[42] With higher $\chi$ value of the out-of-plane $A_{1g}$ modes (~ 129 cm$^{-1}$ and ~ 152 cm$^{-1}$), than in-plane $E_g$ mode, signifies the stronger coupling of electrons with $A_{1g}$ phonon modes. In general, the temperature variation of the phonon frequency appears mainly because of the thermal and volume expansion, which in turn are related to the crystalline anharmonicity.[43] Bi$_2$GeTe$_4$ has a high degree of anharmonicity, which leads to ultralow thermal conductivity making it a potential thermoelectric material.[18] Usually, the crystalline anharmonicity confines phonons with a shorter phonon life time, $\tau$. Therefore, using FWHM of the Raman peaks, the phonon life-time, $\tau_i = \dfrac{1}{2\pi FWHM_i}$ , has been estimated and found to be ranging from 0.33 - 1.4 $ps$.[41] Furthermore, the anomalous phonon softening of $A_{1g}$ modes at ~129 cm$^{-1}$ and ~152 cm$^{-1}$ has been observed below 10 K as represented (Fig. 5 (c-d)), which have been reported in materials with structural transition or presence of spin-phonon coupling.[44,45] Since, Bi$_2$GeTe$_4$ does not have any structural transformation below room temperature, the possible explanation for the anomalous phonon softening at temperatures below 10 K is associated with electron-phonon coupling.[18]

In summary, we showed the involvement of phonons and electron-phonon coupling for topological behavior in single crystal Bi$_2$GeTe$_4$. The existence of topological surface states has been realized from presence of the single Dirac cone in the surface calculations and existence of WAL with large phase coherence length, $l_\phi$ ~ 143 nm. The out-of- plane $A_{1g}$ modes, involving stretching of Bi and Te, have larger anharmonicity than the in-plane $E_g$ modes signifying a stronger coupling of $A_{1g}$ phonons with electrons. Our results provide a subtle way of studying topological insulating properties and involvement of the vibrations of high SOC elements (stretching of Bi-Te bonds) of $Bi_2GeTe_4$.

Acknowledgements: NKS would like to thank CSIR for providing fellowship. AS would like to acknowledge funding support from SERB (Grant No. CRG/2018/002197) and DST India for Indo-Sweden bilateral grant (DST/INT/SWD/VR/P-18/2019). Funding is acknowledged from the Swedish Research Council (VR) under Project No. 2016-03365 and Sweden-India exchange grant 2018-07070, the Knut and Alice Wallenberg Foundation through the Wallenberg Academy Fellows program (grant no. KAW 2020.0196) and support of the Electron Microscopy Laboratory at Linköping University, the Swedish Government Strategic Research Area in Materials Science on Functional Materials at Linköping University (Faculty Grant SFO-Mat-LiU No. 2009 00971) and the Swedish Energy Agency under project 46519-1. P.O.Å.P. acknowledge support from the Swedish Foundation for Strategic Research (SSF) for the Research Infrastructure Fellow program no. RIF 14-0074.

References:

[1]     J. E. Moore, Nature **464**, 194 (2010).

[2]     P. Liu, J. R. Williams, and J. J. Cha, Nature Reviews Materials **4**, 479 (2019).

[3]     M. Z. Hasan and C. L. Kane, Reviews of Modern Physics **82**, 3045 (2010).

[4]     L. Fu, C. L. Kane, and E. J. Mele, Physical Review Letters **98**, 106803 (2007).

[5]     M. Neupane *et al.*, Physical Review B **85**, 235406 (2012).

[6]     T. Trivedi, S. Sonde, H. C. P. Movva, and S. K. Banerjee, Journal of Applied Physics **119** (2016).

[7]     A. Sekine and K. Nomura, Journal of Applied Physics **129**, 141101 (2021).

[8]     M. König, S. Wiedmann, C. Brüne, A. Roth, H. Buhmann, L. W. Molenkamp, X.-L. Qi, and S.-C. Zhang, Science **318**, 766 (2007).

[9]     Y. L. Chen *et al.*, Science **325**, 178 (2009).

[10]    G. M. Stephen, O. A. Vail, J. Lu, W. A. Beck, P. J. Taylor, and A. L. Friedman, Scientific Reports **10**, 4845 (2020).

[11]    R. J. Cava, H. Ji, M. K. Fuccillo, Q. D. Gibson, and Y. S. Hor, Journal of Materials Chemistry C **1**, 3176 (2013).

[12]    L. Fu and C. L. Kane, Physical Review B **76**, 045302 (2007).

[13]    K. Okamoto *et al.*, Physical Review B **86**, 195304 (2012).

[14]    O. G. Karpinsky, L. E. Shelimova, M. A. Kretova, and J. P. Fleurial, Journal of Alloys and Compounds **265**, 170 (1998).

[15]    A. Soni, Z. Yanyuan, Y. Ligen, M. K. K. Aik, M. S. Dresselhaus, and Q. Xiong, Nano Letters **12**, 1203 (2012).

[16]    D. Wu, L. Xie, X. Xu, and J. He, Advanced Functional Materials **29**, 1806613 (2019).

[17]    N. K. Singh, J. Pandey, S. Acharya, and A. Soni, Journal of Alloys and Compounds **746**, 350 (2018).

[18]    N. K. Singh and A. Soni, Applied Physics Letters **117**, 123901 (2020).

[19]    (See Supplemental Material at .... for synthesis and characterization details.).

[20]    K. Momma and F. Izumi, Applied Crystallography **44**, 1272 (2011).

[21]    J. Barthel, Ultramicroscopy **193**, 1 (2018).

[22]    W. Setyawan and S. Curtarolo, Computational Materials Science **49**, 299 (2010).

[23]    P. E. Blöchl, Physical Review B **50**, 17953 (1994).

[24]    G. Kresse and J. Furthmüller, Physical Review B **54**, 11169 (1996).

[25]    G. Kresse and J. Furthmüller, Computational Materials Science **6**, 15 (1996).

[26]    G. Kresse and D. Joubert, Physical Review B **59**, 1758 (1999).

[27]    J. P. Heremans, Nature Physics **11**, 990 (2015).

[28]    Takao Kino, Toshiro Endo, and S. Kawata, Journal of the Physical Society of Japan **36**, 698 (1974).

[29]    W. T. Pawlewicz, J. A. Rayne, and R. W. Ure, Physics Letters A **48**, 391 (1974).

[30]    M. Busch, O. Chiatti, S. Pezzini, S. Wiedmann, J. Sánchez-Barriga, O. Rader, L. V. Yashina, and S. F. Fischer, Scientific Reports **8**, 485 (2018).

[31]    Shama, R. K. Gopal, G. Sheet, and Y. Singh, Scientific Reports **11**, 12618 (2021).

[32]    S. Hikami, A. I. Larkin, and Y. Nagaoka, Progress of Theoretical Physics **63**, 707 (1980).

[33]    P. Sahu, J.-Y. Chen, J. C. Myers, and J.-P. Wang, Applied Physics Letters **112**, 122402 (2018).

[34]    J. Chen, X. Y. He, K. H. Wu, Z. Q. Ji, L. Lu, J. R. Shi, J. H. Smet, and Y. Q. Li, Physical Review B **83**, 241304 (2011).

[35]    Z. Wang, L. Yang, X. Zhao, Z. Zhang, and X. P. A. Gao, Nano Research **8**, 2963 (2015).

[36]    S.-P. Chiu and J.-J. Lin, Physical Review B **87**, 035122 (2013).

[37]    K. M. F. Shahil, M. Z. Hossain, V. Goyal, and A. A. Balandin, Journal of Applied Physics **111**, 054305 (2012).

[38]    J. A. Sans *et al.*, Inorganic Chemistry **59**, 9900 (2020).

[39]    S. Baroni, S. de Gironcoli, A. Dal Corso, and P. Giannozzi, Reviews of Modern Physics **73**, 515 (2001).

[40]    A. Togo and I. Tanaka, Scripta Materialia **108**, 1 (2015).

[41]  J. Pandey, S. Mukherjee, D. Rawat, S. Athar, K. S. Rana, R. C. Mallik, and A. Soni, ACS Applied Energy Materials **3**, 2175 (2020).

[42]  N. Peimyoo, J. Shang, Y. Weihuang, Y. Wang, C. Cong, and T. Yu, Nano Res. **8**, 1 (2014).

[43]  P. S. Peercy and B. Morosin, Physical Review B **7**, 2779 (1973).

[44]  X.-B. Chen, N. T. M. Hien, K. Han, J. C. Sur, N. H. Sung, B. K. Cho, and I.-S. Yang, Journal of Applied Physics **114**, 013912 (2013).

[45]  J. Vermette, S. Jandl, and M. M. Gospodinov, Journal of Physics: Condensed Matter **20**, 425219 (2008).

# Supplemental Material

# Electron-Phonon Coupling and Quantum Correction to Topological Magnetoconductivity in Bi₂GeTe₄

Niraj Kumar Singh,[1] Divya Rawat,[1] Dibyendu Dey,[2] Anna Elsukova,[3] Per O. Å. Persson,[3] Per Eklund,[3] A. Taraphder[4] and Ajay Soni[1]*

[1]School of Basic Sciences, Indian Institute of Technology Mandi, Mandi 175075, Himachal Pradesh, India

[2]Department of Physics and Astronomy, University of Maine, Maine 04469, USA

[3]Thin Film Physics Division, Department of Physics Chemistry and Biology (IFM), Linköping University, Linköping, SE-58183, Sweden

[4]Centre for Theoretical Studies, Department of Physics, Indian Institute of Technology Kharagpur, Kharagpur 721302, India

*Author to whom correspondence should be addressed: ajay@iitmandi.ac.in

In this supplemental file we provide details of the sample synthesis, characterization techniques and selected data complementing the main text.

### (a) Synthesis and characterization details.

Single crystals of Bi₂GeTe₄ were grown using conventional melt grown method using high purity chunks of Bi (99.99 %), Ge (99.999%) and Te (99.99%). The precursors were flame-sealed in a quartz ampoule under vacuum (base pressure below ~ $10^{-5}$ mbar) and subjected to heat treatment at 1073 K with a ramping rate of 4 K min$^{-1}$. The tube was kept at 1073 K for 12 h and subsequently cooled slowly to 673 K in the next 48 h, followed by ambient cooling down to room temperature. The obtained ingot, with shiny metallic luster, was mechanically cleaved and exfoliated from top to remove any possible surface contamination. The exfoliated flake was used for structural and magnetotransport characterizations. X-ray diffraction (XRD) for determining phase purity and crystallographic analysis was carried out using a Rigaku Smartlab X-ray diffractometer with $CuK_\alpha$ radiation. Morphological studies and elemental distribution mapping on freshly cleaved flakes were carried out by employing field-emission scanning electron microscope (FESEM, JFEI, USA, Nova Nano SEM 450). Transmission electron microscopy (TEM) including high-resolution scanning TEM (HR-STEM) imaging was performed in the double-corrected FEI Titan³ 60-300, operated at 300 kV. A cross-sectional TEM lamella was prepared using focused ion beam method (Carl Zeiss Cross-Beam 1540 EsB system). The simulated diffraction pattern was generated using Single Crystal™: a single-crystal diffraction program (Crystal Maker Software Ltd, Oxford, England). The crystal structure was visualized using VESTA software. HR-STEM image simulations were performed using Dr. Probe software.[1, 2] The phonon dynamics and the vibrational properties were studied using Jobin-Yvon Horiba LabRAM HR evolution Raman spectrometer using a 532 nm excitation laser. The spectra were recorded using 1800 grooves/mm grating and ultralow frequency filters to access low-frequency Raman modes. All the spectra were fitted with multiple Lorentzian fits in order to evaluate peak position, full width at half maxima (FWHM), and integrated area under the curve. Temperature dependent Raman measurements were performed in the range of 3 – 300 K using a Montana cryostation. The low temperature (2 – 300 K) transport properties were measured using a commercial physical properties measurement system (PPMS, Quantum Design, Dynacool).

**(b) Density functional theory computational details.**

Density functional theory (DFT) calculations were performed using a plane-wave basis set with a kinetic energy cutoff of 400 eV and projector augmented-wave potentials, as implemented in the Vienna Ab initio Simulation Package (VASP).[3-6] For the exchange-correlation functional, local density approximations (LDA) was used. Bulk calculations were performed using the rhombohedral unit cell of $Bi_2GeTe_4$, whereas a supercell made up of five stacked seven-atomic-layer slabs with a vacuum of 20 Å along the (111) direction was constructed to perform surface calculations. The reciprocal space integration was carried out with a $\Gamma$-centered k-mesh of $8\times8\times8$ for the bulk and $12\times12\times1$ for the surface calculations. In order to get the precise band dispersions, spin-orbit coupling (SOC) was included in our calculations. During structural relaxations performed within LDA, positions of the ions were relaxed until the Hellman-Feynman forces became less than 0.001 eV/Å. Phonons were calculated using density functional perturbation theory (DFPT) as implemented in the PHONOPY.[7, 8]

**(c) Rietveld refinement analysis.**

The as-synthesized $Bi_2GeTe_4$ ingot was ground into fine powder for XRD analysis. The phase purity of $Bi_2GeTe_4$ sample has been confirmed by Rietveld refinements of the powder XRD pattern. The Fig. S1 shows the Rietveld refined XRD data. A detailed account of the extracted parameters is presented in Table I.

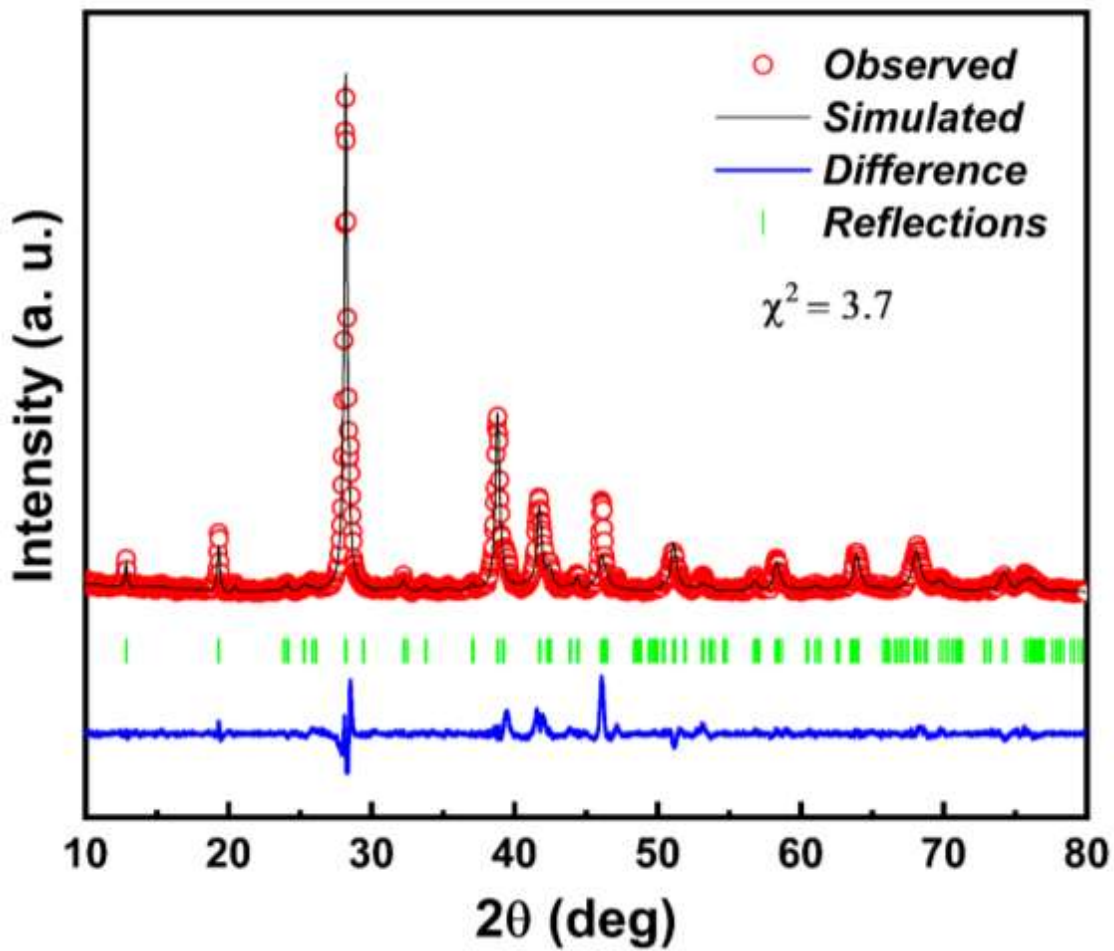

*FIG. S1 Rietveld refinement of XRD pattern of powdered Bi₂GeTe₄.*

Table S1. The extracted lattice parameters from Rietveld refinement and goodness of fit. Here, $\chi^2$ represents the goodness of fit, whereas $a$, $b$ and $c$ are the lattice parameters. The parameters $R_p$, $R_{wp}$ and $R_e$ denotes the profile factor, weighted profile factor and expected R factor.

| $\chi^2$ | $a$ (Å) | $b$ (Å) | $c$ (Å) | $V(Å^3)$ | $R_p$ | $R_{wp}$ | $R_e$ |
|---|---|---|---|---|---|---|---|
| 3.63 | 4.3283 | 4.3283 | 41.3578 | 670.9873 | 20.8 | 24.7 | 13 |

**(d) FESEM and elemental distribution mapping.**

The field emission-scanning electron microscopy (FESEM) image of cross-section of Bi$_2$GeTe$_4$ single crystalline flake and the distribution of constituent elements are shown in Fig. S2. The images showcase the homogeneous distribution of elements.

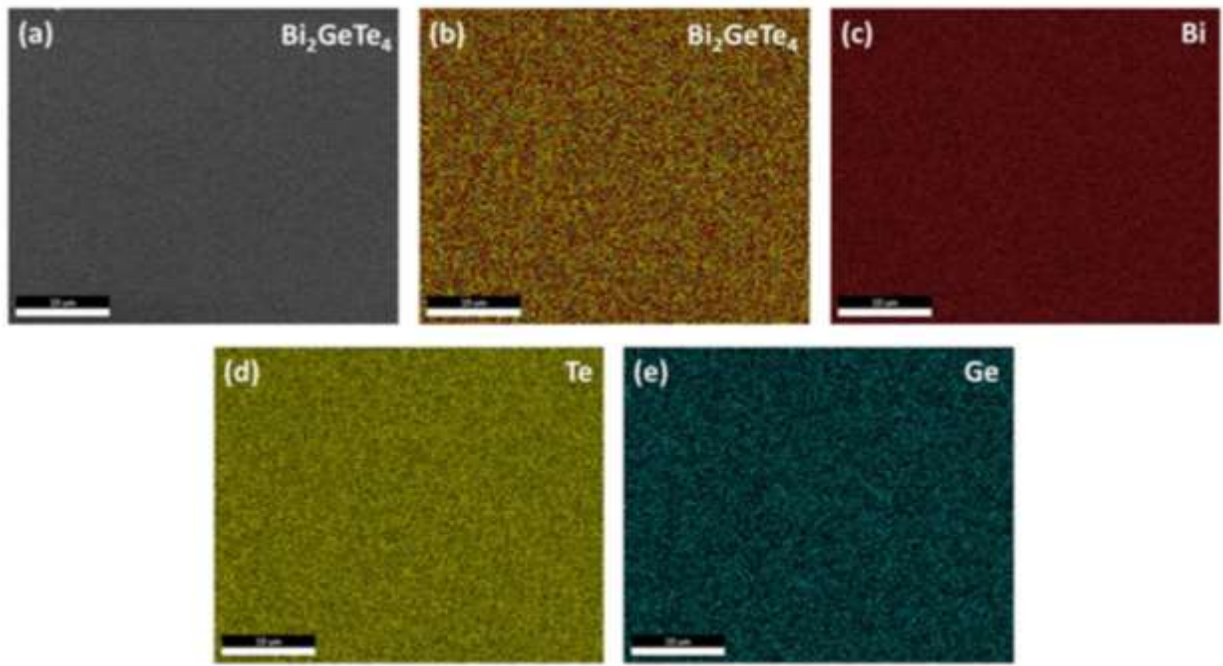

FIG. S2 (a) Top flat surface of the flake and (b-e) Elemental distribution mapping on the surface of the flake.

**(e) Bulk electronic band structure calculation without SOC.**

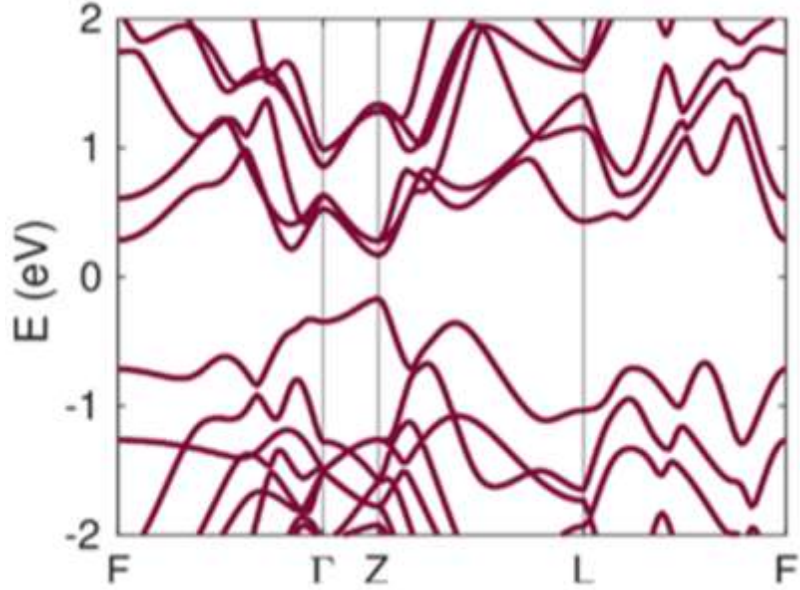

FIG. S3 Band structure of bulk Bi$_2$GeTe$_4$ without SOC showing an over estimated large electronic band gap.

**(f) Table S3: Comparison of experimentally observed and theoretically calculated Raman active modes.**

| Modes | $\omega_{th}$ (cm$^{-1}$) | $\omega_{exp}$ (cm$^{-1}$) | |
|---|---|---|---|
| | | at 300K | at 4K |
| E$_g$ | 33.6 | 28.2 | 29.8 |
| A$_{1g}$ | 51.2 | 48.3 | 50.4 |
| E$_g$ | 103.2 | - | 94.6 |
| E$_g$ | 110.9 | 101.3 | 105.7 |
| A$_{1g}$ | 116.4 | 129 | 136.3 |
| A$_{1g}$ | 146.1 | 152 | 159.8 |

**(g) Analysis of Temperature dependence of Raman modes.**

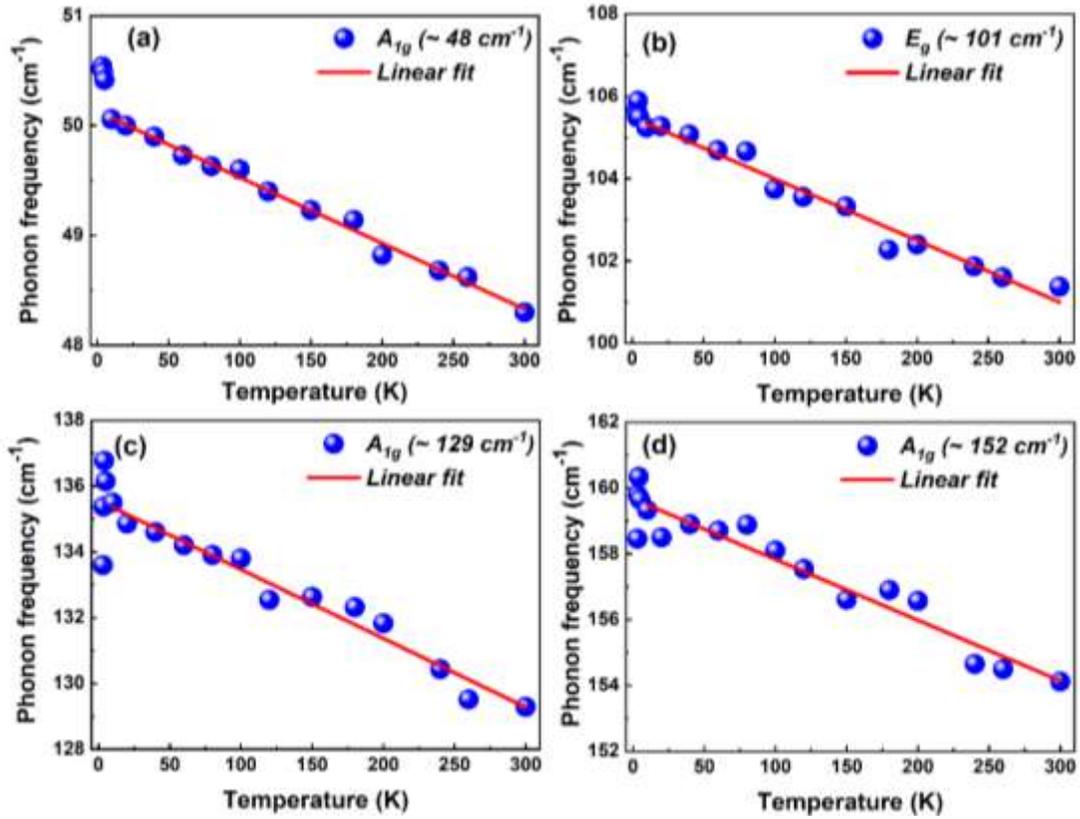

FIG. S4 Temperature dependence of the phonon frequency of the various Raman modes.

References:

1.    Momma, K. and F. Izumi, *VESTA 3 for three-dimensional visualization of crystal, volumetric and morphology data.* Applied Crystallography 2011. **44**(6): p. 1272-1276.

2.    Barthel, J., *Dr. Probe: A software for high-resolution STEM image simulation.* Ultramicroscopy, 2018. **193**: p. 1-11.

3.    Blöchl, P.E., *Projector augmented-wave method.* Physical Review B, 1994. **50**(24): p. 17953-17979.

4.    Kresse, G. and D. Joubert, *From ultrasoft pseudopotentials to the projector augmented-wave method.* Physical Review B, 1999. **59**(3): p. 1758-1775.

5.      Kresse, G. and J. Furthmüller, *Efficient iterative schemes for ab initio total-energy calculations using a plane-wave basis set.* Physical Review B, 1996. **54**(16): p. 11169-11186.

6.      Kresse, G. and J. Furthmüller, *Efficiency of ab-initio total energy calculations for metals and semiconductors using a plane-wave basis set.* Computational Materials Science, 1996. **6**(1): p. 15-50.

7.      Baroni, S., et al., *Phonons and related crystal properties from density-functional perturbation theory.* Reviews of Modern Physics, 2001. **73**(2): p. 515-562.

8.      Togo, A. and I. Tanaka, *First principles phonon calculations in materials science.* Scripta Materialia, 2015. **108**: p. 1-5.